\documentclass[twocolumn,prd,aps,floats,showpacs]{revtex4}
\def\DESepsf(#1 width #2){\epsfxsize=#2 \epsfbox{#1}}
\usepackage{graphicx}
\usepackage{color}
\usepackage{amsmath}
\def\bmatrix{\left[\begin{array}}
\def\ematrix{\end{array}\right]}

\begin{document}

%

\let\a=\alpha      \let\b=\beta       \let\c=\chi        \let\d=\delta
\let\e=\varepsilon \let\f=\varphi     \let\g=\gamma      \let\h=\eta
\let\k=\kappa      \let\l=\lambda     \let\m=\mu
\let\o=\omega      \let\r=\varrho     \let\s=\sigma
\let\t=\tau        \let\th=\vartheta  \let\y=\upsilon    \let\x=\xi
\let\z=\zeta       \let\io=\iota      \let\vp=\varpi     \let\ro=\rho
\let\ph=\phi       \let\ep=\epsilon   \let\te=\theta
\let\n=\nu
\let\D=\Delta   \let\F=\Phi    \let\G=\Gamma  \let\L=\Lambda
\let\O=\Omega   \let\P=\Pi     \let\Ps=\Psi   \let\Si=\Sigma
\let\Th=\Theta  \let\X=\Xi     \let\Y=\Upsilon

%

%

\def\cA{{\cal A}}                \def\cB{{\cal B}}
\def\cC{{\cal C}}                \def\cD{{\cal D}}
\def\cE{{\cal E}}                \def\cF{{\cal F}}
\def\cG{{\cal G}}                \def\cH{{\cal H}}
\def\cI{{\cal I}}                \def\cJ{{\cal J}}
\def\cK{{\cal K}}                \def\cL{{\cal L}}
\def\cM{{\cal M}}                \def\cN{{\cal N}}
\def\cO{{\cal O}}                \def\cP{{\cal P}}
\def\cQ{{\cal Q}}                \def\cR{{\cal R}}
\def\cS{{\cal S}}                \def\cT{{\cal T}}
\def\cU{{\cal U}}                \def\cV{{\cal V}}
\def\cW{{\cal W}}                \def\cX{{\cal X}}
\def\cY{{\cal Y}}                \def\cZ{{\cal Z}}
%

\newcommand{\Ns}{N\hspace{-4.7mm}\not\hspace{2.7mm}}
\newcommand{\qs}{q\hspace{-3.7mm}\not\hspace{3.4mm}}
\newcommand{\ps}{p\hspace{-3.3mm}\not\hspace{1.2mm}}
\newcommand{\ks}{k\hspace{-3.3mm}\not\hspace{1.2mm}}
\newcommand{\des}{\partial\hspace{-4.mm}\not\hspace{2.5mm}}
\newcommand{\desco}{D\hspace{-4mm}\not\hspace{2mm}}
\renewcommand{\figurename}{Fig.}


%
\title{\boldmath
Fourth Generation Leptons and Muon $g-2$}
\vfill
\author{Wei-Shu Hou}
\author{Fei-Fan Lee}
\author{Chien-Yi Ma}
\affiliation{
 Department of Physics,
 National Taiwan University,
 Taipei, Taiwan 10617, R.O.C.
}

\date{\today}
%
%
%
\begin{abstract}
We consider the contributions to $g_\mu-2$ from fourth generation
heavy neutral and charged leptons, $N$ and $E$, at the one-loop
level. Diagrammatically, there are two types of contributions:
boson-boson-$N$, and $E$-$E$-boson in the loop diagram. In
general, the effect from $N$ is suppressed by off-diagonal lepton
mixing matrix elements. For $E$, we consider flavor changing
neutral couplings arising from various New Physics models, which
are stringently constrained by $\mu\to e\gamma$. We assess how the
existence of a fourth generation would affect these New Physics
models.
\end{abstract}
%
\maketitle

\pagestyle{plain}

\section{Introduction}

The fourth generation has been viewed as out of favor~\cite{PDG}
since a long time, because of electroweak precision tests (EWPrT),
and neutrino counting on the $Z$ peak. However, the severeness of
the $S$ parameter constraint from EWPrT has been questioned
recently~\cite{KPST07}, while we know that the neutrino sector is
much richer than originally thought because of neutrino
oscillations. With the advent of the LHC, we now have a machine
which can discover or rule out the 4th generation by direct
search, once and for all~\cite{AH06}. Currently, the Tevatron has
set stringent limits~\cite{tpCDF08} on $t'$ via $t'\to qW$ search.

It was recently pointed out~\cite{Hou08} that the existence of a
4th generation could have implications for the baryon asymmetry of
the Universe (BAU). By shifting the Jarlskog
invariant~\cite{Jarlskog} for $CP$ violation (CPV) of the 3
generation Standard Model (SM3) by one generation, i.e. from 1-2-3
to 2-3-4 quarks, one gains by more than $10^{13}$ in effective
CPV, and may be sufficient for BAU\,! Recent developments in CPV
studies at the B factories~\cite{BelleNature} and the
Tevatron~\cite{CDFsin2betas} suggest the 4th generation could be
behind some hints for New Physics in $b\to s$ transitions.
From a different perspective, whether from effective 4-fermion
interactions~\cite{Holdom}, or from holographic extra dimension
considerations~\cite{Burdman} (the two are complementary), there
are also recent interest in very heavy 4th generation quarks,
where their heaviness could be responsible for inducing
electroweak symmetry breaking itself.

With renewed interest in the existence of a sequential 4th
generation, and with experimental discovery or refutation expected
at the LHC in due time, we turn to the lepton sector. Our goal is
modest: if a 4th generation exists, what are the implications for
the most prominent probes with charged leptons, i.e. muon $g-2$,
$\mu\to e\gamma$, and $\tau \to \ell\gamma$?

The difference between the experimental value and the SM3
prediction of muon $g-2$ has been around for some time
now~\cite{Stoeckinger}. That is,
\begin{eqnarray}
 a_\mu^{\mathrm{exp}}-a_\mu^{\mathrm{SM}}=295(88)\times10^{-11},
 \label{difference}
\end{eqnarray}
where $a_\mu\equiv (g_\mu-2)/2$. The difference of over
3.4$\sigma$ has aroused a lot of interest. We also have very
stringent bounds on lepton flavor violating (LFV) rare decays,
such as~\cite{PDG}
\begin{eqnarray}
 {\cal B}(\mu\to e\gamma) < 1.2 \times 10^{-11},
 \label{mutoegam}
\end{eqnarray}
and the $\tau$ decay counterpart~\cite{HFAG}
\begin{eqnarray}
 && {\cal B}(\tau\to e\gamma) < 1.1 \times 10^{-7},
 \label{tautoegam} \\
 && {\cal B}(\tau\to\mu\gamma) < 4.5\times10^{-8},
 \label{tautomugam}
\end{eqnarray}
at 90\% C.L. These limits could be improved further in the near
future. The MEG experiment, a $\mu\to e\gamma$ search experiment
aiming at a sensitivity of $10^{-13}$~\cite{MEG}, has started its
physics run in 2008. The short 2008 run alone is expected to bring
the limit below $10^{-12}$. Though the limits on $\tau \to
\ell\gamma$ from the B factories, Eqs.~(\ref{tautoegam}) and
(\ref{tautomugam}), will soon be limited by B factory statistics,
a Super B Factory upgrade could push down to the $10^{-8}$ region,
which become background limited (for outlook, see
Ref.~\cite{Hayashii}).

\begin{figure}[b!]
  \vskip-0.5cm
  \includegraphics[height=3.4cm]{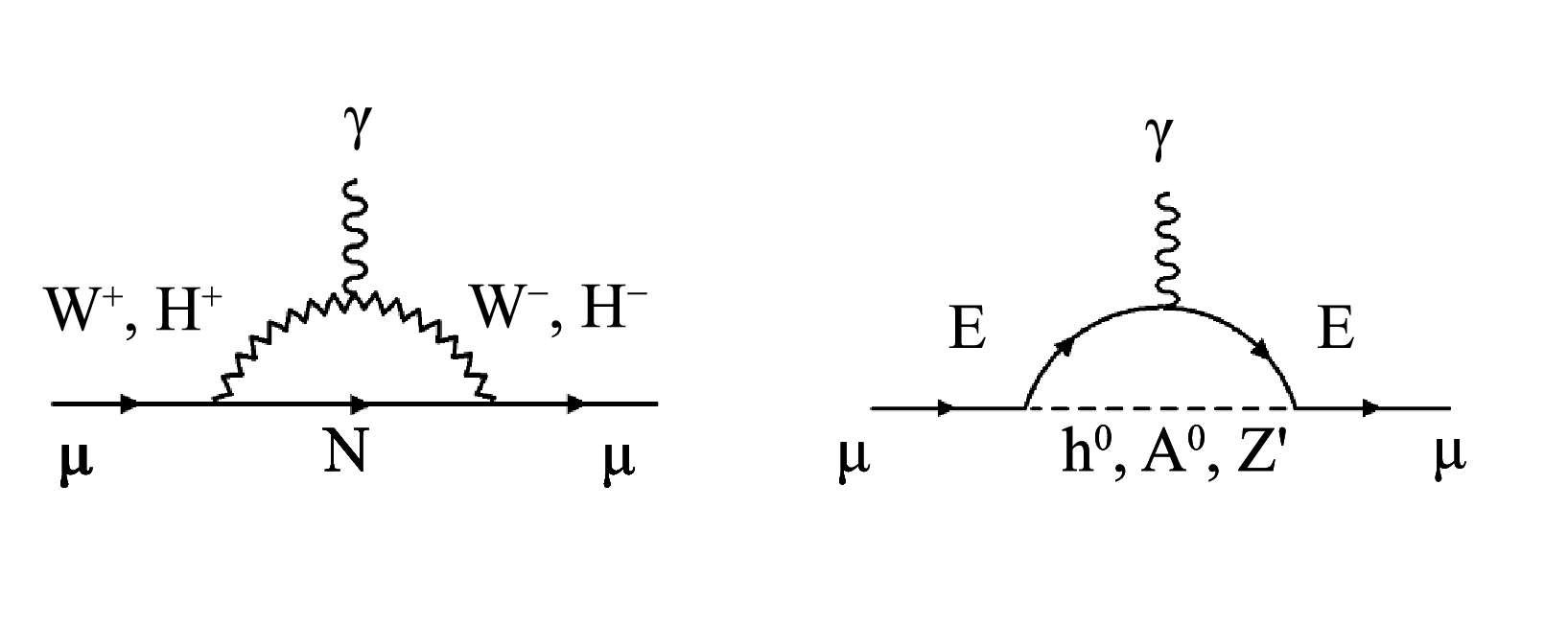}
  \vskip-0.5cm
  \caption{
   (a) Boson-boson-$N$ and
   (b) $E$-$E$-boson loop diagrams.
   }
  \label{fig:41}
\end{figure}

Can the effect of 4th generation leptons show up in the probes of
Eqs.~(\ref{difference})--(\ref{tautomugam})? How would these
processes constrain New Physics models in the presence of a 4th
generation?
In this paper we start our discussion from a diagrammatic point of
view, and in so doing, correct some errors in the literature.

The 4th generation neutral lepton $N$ can enter the loop with
charged vector boson $W^{\pm}$ or scalar boson $H^{\pm}$, which we
plot for $\mu\mu\gamma$ coupling in Fig.~1(a) for illustration.
The diagrams for $\mu \to e \gamma$ and $\tau \to \ell \gamma$ are
quite similar. The $W^+W^-N$ loop is controlled by the lepton
mixing matrix elements, while the $H^+H^-N$ loop may become
important because of $m_N$. The charged lepton $E$ can enter the
loop with neutral scalar and pseudo scalar bosons $h^0$ and $A^0$,
or a neutral vector boson $Z'$, as illustrated in Fig.~1(b).
However, these neutral bosons would need to have flavor changing
neutral couplings (FCNC, absent in SM) to be relevant. But then
they will face stringent constraints from $\mu \to e \gamma$ and
$\tau \to \ell\gamma$. Therefore, if the 4th generation exists,
Eqs.~(\ref{difference})--(\ref{tautomugam}) will constrain New
Physics models.

In the next section we first discuss the contributions involving
neutral lepton $N$. In Sec. III we discuss the contributions
involving charged lepton $E$. In Sec. IV we compare with the
minimal supersymmetric SM (MSSM). A summary is given in Sec. V.

\section{Effects of Neutral Lepton $N$}

The 4th generation neutral lepton $N$ enters the one-loop diagram
for muon $g-2$ illustrated in Fig. 1(a). The boson can be the
charged vector boson $W^{\pm}$, or charged scalar boson $H^{\pm}$,
which we discuss separately.

\subsection{\boldmath $W^{+}W^{-}N$ Loop Contribution}

With $\nu_\mu$ instead of $N$, this is the only contribution
within SM. The contribution from a fourth generation lepton $N$
has been considered before~\cite{HuoFeng,Lynch1}. We find
\begin{eqnarray}
 a_\mu(W^{+}W^{-}N)
 &=&\frac{G_{\mathrm{F}}m_\mu^2}{4\sqrt{2}\pi^2}\,|V_{N\mu}|^2\, F(x),
 \label{WWN}
\end{eqnarray}
where $x = m^2_{N}/M^2_{W}$, $V_{N\mu}$ is the lepton mixing
matrix element, and
\begin{eqnarray}
 F(x)
 &=&\int_{0}^{1}du\,\frac{u(1-u)(2-u)x+2u^{2}(u+1)}
 {(1-u)x+u} \nonumber\\
 &=&\frac{3x^3\log x}{(x-1)^4} +
    \frac{4x^3-45x^2+33x-10}{6(x-1)^3}.\ \
 \label{fx1}
\end{eqnarray}

\begin{figure}[t!]
 \centering
  \includegraphics[height=4.5cm]{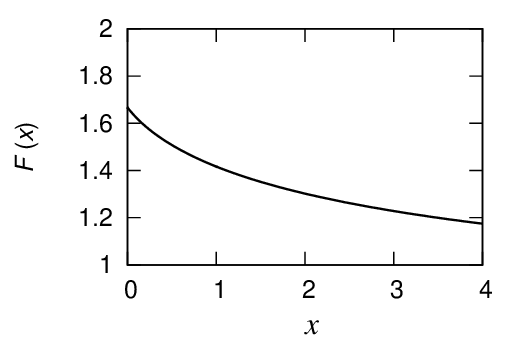}
  \caption{
   Loop function $F(x)$ of Eq.~(\ref{WWN}), vs $x = m_N^2/M_W^2$.
  }
 \label{fig:44}
\end{figure}
%
%
%

We depict $F(x)$ versus $x$ in Fig.~\ref{fig:44}. We see that
$F(x)$, an Inami--Lim~\cite{IL81} loop function, is well-behaved
and bounded, with $F(1)=17/12$. However, this does not seem to be
correctly rendered in Refs.~\cite{HuoFeng} and \cite{Lynch1}.
In Ref.~\cite{HuoFeng}, though bounded, $F(x)$ was not correctly
evaluated. Furthermore, the authors found strong enhancement near
$x \sim 1$, which was used to put a bound on the fourth neutral
lepton mass. However, from the integral in Eq.~(\ref{fx1}) and the
functional form of $F(x)$, it should be clear that there is no
enhancement near $x = 1$ (i.e. $m_N \sim M_W$).
In Ref.~\cite{Lynch1}, which is a study note for
Ref.~\cite{Lynch}, the form of $F(x)$ is again incorrect but still
bounded. The author claimed that $F(x)$ had a singularity at $x =
1$, and attributed this to the zero width approximation of the $W$
boson propagator. Again, we see from Eq.~(\ref{fx1}), that there
is no singularity for any $x$, and in any case, $\Gamma_W$ should
be irrelevant for such low scale processes. Our result therefore
corrects some errors in the literature~\cite{Biggio}.

As we have already stated, if we replace $N$ by $\nu_\mu$, we
should recover the SM contribution. Using $F(0)=5/3$ in
Eq.~(\ref{WWN}), together with $V_{\nu_\mu\mu} \cong~1$, we get
$a_\mu^{\mathrm{SM}}(W^{+}W^{-}\nu_\mu) =
5G_{\mathrm{F}}m_\mu^2/12\sqrt{2}\pi^2$. Furthermore, we find
$a_\mu^{\mathrm{SM}}(\mu\mu Z) =
-(G_{\mathrm{F}}m_\mu^2/6\sqrt{2}\pi^2)(1+2\sin^2\theta_W-4\sin^4\theta_W)$,
where $\sin^2\theta_W=0.23120$. Combining the two together, and
using
\begin{eqnarray}
 \frac{G_F m_\mu^2}{4\sqrt{2}\pi^2} &\simeq& 233 \times 10^{-11},
 \label{fac}
\end{eqnarray}
we get $a_\mu^{\mathrm{SM}}(\text{1-loop Electroweak})=195 \times
10^{-11}$, which is consistent with Ref. \cite{Marciano}.

The SM exercise indicates that the 4th generation neutral lepton
contribution has the right order of magnitude to contribute to
Eq.~(\ref{difference}). However, as seen from Fig.~\ref{fig:44},
the effect actually drops a bit from the massless $\nu_\mu$ result
of SM as the mass of $N$ becomes heavier. Furthermore, it is
multiplied by the suppression factor $\vert V_{N\mu}\vert^2$. From
$m_N \gtrsim 90$ GeV~\cite{PDG}, hence $F(x) \lesssim 1.4$, we see
that $|V_{N\mu}|$ needs to be 0.7 or higher to reach within
$2\sigma$ of Eq.~(\ref{difference}). Considering the stringent
constraint from Eq.~(\ref{mutoegam}), however, this is clearly
unrealistic. We conclude that the difference of
Eq.~(\ref{difference}) cannot come from the addition of a 4th
neutral lepton $N$.

\subsection{\boldmath $H^+ H^- N$ Loop Contribution}

It is unusual to consider both a 4th neutral lepton $N$ together
with charged Higgs $H^+$. But since $W^+W^-N$ contribution is
insufficient for Eq.~(\ref{difference}), we consider replacing
$W^{+}$ by the charged Higgs $H^{+}$. This is the
Two-Higgs-Doublet-Model (2HDM) with 4th generation leptons. It is
of interest to check whether one could gain from large $\tan\beta$
enhancement.

For 2HDM-II (which occurs for MSSM), where up and down type quarks receive
masses from different Higgs doublets, we find
\begin{eqnarray}
 &&a_\mu^{\rm 2HDM-II}(H^+ H^- N) \nonumber\\
 &=&-\frac{G_{F}\,m_{\mu}^2}{4\sqrt{2}\pi^2} |V_{N\mu}|^{2}\,[f_{H^+}(x)\nonumber\\
 &&\hspace{1cm}+g_{H^+}(x)\cot^2\beta
    +x_\mu\,q_{H^+}(x)\tan^2\beta],\ \
 \label{HHN}
\end{eqnarray}
where $x=m_{N}^2/M_{{H}^+}^2$ and $x_\mu=m_{\mu}^2/M_{H^+}^2$.
The loop functions in Eq.~(\ref{HHN}) are
\begin{eqnarray}
  f_{{H}^+}(x)&=&
   \int_{0}^{1}du\,
   \frac{2u(1-u)x}{(1-u)\,x+u\,}\nonumber\\
   &=&-\frac{2x^2\log x}{(x-1)^3}+\frac{x(x+1)}{(x-1)^2},\ \
 \label{fHx} \\
%
%
  g_{{H}^+}(x)&=&
   \int_{0}^{1}du\,
   \frac{u^2(1-u)x}{(1-u)\,x+u\,}\nonumber\\
   &=&-\frac{\,{x}^{3}\,\log\,x}{(x-1)^4}+
   \frac{x(2x^2+5x-1)}{6(x-1)^3},\ \
 \label{gHx} \\
%
  q_{{H}^+}(x)&=&
   \int_{0}^{1}du\,
   \frac{u^2(1-u)}{(1-u)\,x+u\,}\nonumber\\
   &=&-\frac{{x}^{2}\,\log~x }
   {(x-1)^4}+\frac{2\,{x}^{2}+5\,x-1}{6(x-1)^3}.\ \
 \label{qHx}
\end{eqnarray}

\begin{figure}[t!]
 \centering
  \includegraphics[height=4.5cm]{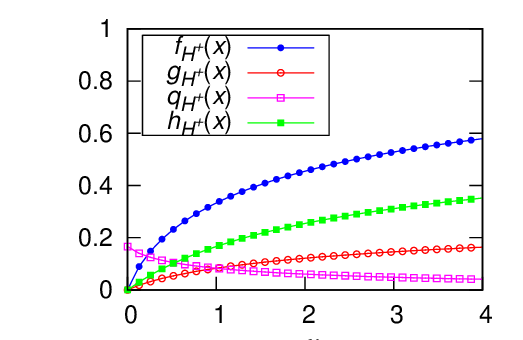}
  \caption{
   Loop functions $f_{H^+}(x)$, $g_{H^+}(x)$, $q_{H^+}(x)$
   of Eq.~(\ref{HHN}), and $h_{H^+}(x)$, $q_{H^+}(x)$ of
   Eq.~(\ref{HHN2HDMI}) vs $x=m_N^2/M_{H^+}^2$.
   }
\end{figure}

We plot $f_{H^+}(x)$, $g_{H^+}(x)$ and $q_{H^+}(x)$ in Fig.~3. The
$q_{H^+}(x)$ term in Eq. (\ref{HHN}) can actually be safely
ignored, because one would need extremely large values of
$\tan\beta$ to overcome the extremely small $x_\mu$.
However, we give a complete expression to check a previous result
in Ref.~\cite{Dedes_Haber} for 3 generations. If we replace $N$ be
$\nu_\mu$, one has $f_{H^+}(0)=0$, $g_{H^+}(0)=0$ and
$q_{H^+}(0)=1/6$, and Eq.~(\ref{HHN}) becomes
$a_\mu^{\mathrm{2HDM-II}}(H^+H^-\nu_\mu)=-(G_F
m_\mu^2/4\sqrt{2})x_\mu\,\tan^2\beta/6$.

For the first term of Eq. (\ref{HHN}), for $1 \lesssim x \lesssim
10$ we have $0.4 \lesssim f_{H^+}(x) \lesssim 0.8$, which is not
particularly small. But because of the general $|V_{N\mu}|^2$
suppression, an argument similar to the $W^+W^-N$ loop discussion
suggest that this term can not give rise to
Eq.~(\ref{difference}). It is interesting that, because $N$ has
isospin $+1/2$, large $\cot\beta$ could lead to enhancement. If we
take $|V_{N\mu} \cot\beta|^2$ to be order 1 in the large
$\cot\beta$ limit, and if $m_N$ is large compared to $m_{H^+}$, it
could generate a finite contribution. But this contribution is
{\it negative}, hence it is in the wrong direction for $\Delta
a_\mu$ of Eq.~(\ref{difference}). Furthermore, in the 2HDM-II (or
MSSM), the $t\bar{t}H^0(h^0)$ coupling relative to its SM value,
$m_t/v$, is given by $\cos\alpha/\sin\beta$
($\sin\alpha/\sin\beta$). Large $\cot\beta$ will make the coupling
strength $|g_{t\bar{t}H^0}|\gg1$ or $|g_{t\bar{t}h^0}|\gg1$ and
become nonperturbative, which is not desirable.

It has already been pointed, however, that there is a sufficient
contribution in MSSM~\cite{Marciano,Moroi,Everett} coming from the
large $\tan\beta$ region. We will give a brief assessment in
Sec.~IV.


For 2HDM-I, where all quarks receive mass from the same Higgs
doublet, we find
  \begin{eqnarray}
 &&a_\mu^{\rm {2HDM-I}}(H^+H^-N)\nonumber\\
 &=&\frac{G_F\,m^2_\mu}{4\sqrt{2}\pi^2}|V_{N\mu}|^2\cot^2\beta
 [h_{H^+}(x)-x_\mu\,q_{H^+}(x)],\label{HHN2HDMI}
  \end{eqnarray}
with $x$ and $x_\mu$ as before, and
  \begin{eqnarray}
  h_{H^+}(x)
 &=&\int_0^1du\,\frac{u(1-u)(2-u)x}{(1-u)x+u}\nonumber\\
 &=& - \frac{x^2(x-2)\log\,x}{(x-1)^4}
     + \frac{x(4\,{x}^{2}-5\,x-5)}{6(x-1)^3},\ \
  \label{hHx}
  \end{eqnarray}
while $q_{H^+}(x)$ is given in Eq. (\ref{qHx}). We plot
$h_{H^+}(x)$ also in Fig.~3.
%
Analogous to 2HDM-II, we have
$a_\mu^{\mathrm{2HDM-I}}(H^+H^-\nu_\mu)=-(G_F
m_\mu^2/4\sqrt{2})x_\mu\,\cot^2\beta/6$, but now everything is
proportional to $\cot^2\beta$.

Similar to the 2HDM-II case, if we take $|V_{N\mu} \cot\beta|^2$
to be order 1, it could generate a finite and {\it positive}
contribution to $\Delta a_\mu$. However, for 2HDM-I, the
$\cot\beta$ enhanced Higgs couplings to $t\bar{t}$ are
non-perturbative at large $\cot\beta$, which leads us to reject
this possibility.

\section{effects of Charged Lepton $E$}

The 4th generation charged lepton $E$ contributes to $a_\mu$ via
$E$-$E$-boson loop diagrams, where the boson has to be neutral,
and can be a scalar ($h^0$), pseudo-scalar ($A^0$), or vector (an
extra $Z'$). But they need to possess flavor changing neutral
couplings (FCNC). We discuss each case separately.

\subsection{\boldmath $E E h^0$, $EEA^0$ Loop Contribution}

There is no $\mu EH^0$ coupling in SM. The same is true for the
2HDM-I and II, and $\mu EH^0$, $\mu Eh^0$ and $\mu EA^0$ couplings
are absent. This is because, by design~\cite{GW77}, the charged
leptons receive mass from just one doublet, and only one matrix
needs to be diagonalized.
In the so-called 2HDM-III, this restriction is softened, and there
exist two matrices $\eta^{e(\nu)}$ and $\xi^{e(\nu)}$
simultaneously for each lepton type.
  Note that in this model, by redefining
  $\phi_1$, $\phi_2$ and $\eta$, $\xi$ simultaneously,
  which still leaves the Lagrangian invariant,
  we may assume $\langle\phi_1^0\rangle=v/\sqrt{2}$ and
  $\langle\phi_2^0\rangle=0$ without loss of generality,
  hence $\tan\beta$ is no longer a physical parameter.
  For a detailed analysis, we refer to Ref.~\cite{Diaz}.

To regulate the FCNC in face of stringent constraints, there is
the ansatz suggested by Cheng and Sher~\cite{ChengSher} for the
quark sector, i.e. all $q_iq_jh^0/H^0/A^0$ couplings have the same
form
\begin{eqnarray}
 \Delta_{ij}\frac{\sqrt{m_i m_j}}{v},
 \label{CS}
\end{eqnarray}
where $\Delta_{ij}$ is of $\mathcal{O}(1)$. This scheme can
survive the rather critical constraint of $K^0$--$\bar K^0$
mixing, because $\sqrt{m_d m_s}\,/v$ is extremely small. We extend
it here to the charged lepton sector with 4th generation, although
it may not hold because the lepton mixing pattern seems different
from those of the quarks.

\begin{figure}[t!]
  \centering
  \includegraphics[height=4.5cm]{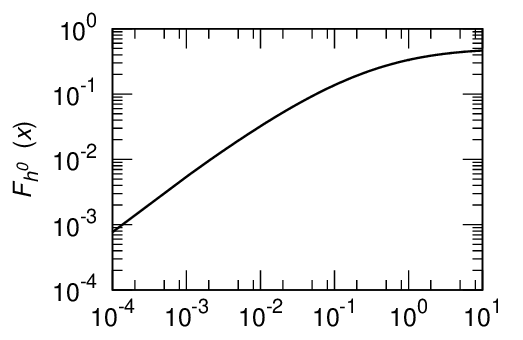}
  \caption{
  Loop function $F_{h^0}(x)$ of Eq. (\ref{EEh}), vs $x=m_E^2/M_{h^0}^2$.
  }
  \label{fig:47}
\end{figure}

Note that CP-even Higgs bosons $H^0, h^0$ give {\it positive}
contributions to $a_\mu$, but CP-odd $A^0$ contributions are {\it
negative}. Considering the positivity of Eq.~(\ref{difference}),
we may assume $A^0$ is very heavy hence can be safely neglected.
For sake of illustration, we set $h^0$ to be the lightest neutral
Higgs, and assume no mixing between $H^0$ and $h^0$. We then find
\begin{eqnarray}
 \Delta a_\mu^{\mathrm{2HDM-III}} \sim 233 \times 10^{-11}\,
 F_{h^0}(x),\label{EEh}
\end{eqnarray}
where $x=m^2_{{E}}/M^2_{{h}^0}$ and we have taken $\Delta_{ij} =
1$, and
\begin{eqnarray}
 F_{h^0}(x)&=&\int_0^1du\,\frac{u^2\,x}
  {u\,x+(1-u)}\nonumber\\
  &=&\frac{x\,\log\,x}{(x-1)^3}+\frac{x(x-3)}{2(x-1)^2}.
  \label{fh0x}
\end{eqnarray}
which is plotted in Fig.~\ref{fig:47}.

There are other loops such as $H^+H^-N$ to be considered, but they
are suppressed by $m_\mu/m_E$ and can be safely neglected. The
suppression factor is $m_\mu/m_E$ rather than $(m_\mu/m_E)^2$
because of the Cheng-Sher coupling enhancement in Eq.~(\ref{CS}).
However, the LFV decay rates in
Eqs.~(\ref{mutoegam})--(\ref{tautomugam}) give very stringent
constraints, and need to be confronted. Here we use the formulas
in Ref.~\cite{PQHung}. Note that because $a_\mu$ and ${\cal
B}(\mu\to e\gamma)$ come from loop diagrams of similar structure,
their formulas are very closely related. After some organization,
we have
 %
 \begin{eqnarray}
 {\cal B}^{\mathrm{2HDM-III}}(\mu\to e\gamma)
 &=&\frac{3\alpha}{2\pi}\frac{m_e}{m_\mu}
 |F_{h^0}(x)|^2\,\nonumber\\
 &=&1.7\times10^{-5} |F_{h^0}(x)|^2\,.\label{meg2HDMIII}
 \end{eqnarray}
 %

Let us consider first the case of $\tau$ in the loop, which is the
leading contribution with 3 generations, and was discussed in
Ref.~\cite{ChangHouKeung}. Eq. (\ref{meg2HDMIII}) becomes
\begin{eqnarray}
 &\,& {\cal B}^{\mathrm{2HDM-III}}(\mu\to e\gamma)
      \vert_{\rm 3\ gen.} \nonumber\\
&=&\frac{3\alpha}{2\pi}\frac{m_e}{m_\mu}
\,\,\left|{\frac{m^2_\tau}{M^2_{h^0}}
\left(\log\,\frac{m_\tau^2}{M^2_{h^0}}+\frac{3}{2}\right)}\right|^2 \,.\label{meg2HDMIIISM3}
\end{eqnarray}
Considering a factor of 2 uncertainty in
Eq.~(\ref{meg2HDMIIISM3}), we still require
$M_{h^0}>138\,\mathrm{GeV}$ in order to survive
Eq.~(\ref{mutoegam}). We note that the MEG experiment can push the
lower bound down to $530\,\mathrm{GeV}$.

Consider now 4 generations. Comparing Eqs.~(\ref{difference}) and
(\ref{EEh}), we have $F_{h^0}(x)=\mathcal{O}(1)$, and it seems
that one could in principle bring about the difference with 4th
generation under Cheng--Sher ansatz. However,
Eqs.~(\ref{mutoegam}) and (\ref{meg2HDMIII}) give $F_{h^0}(x)
\lesssim 10^{-3}$. From Fig.~\ref{fig:47}, we see that $m_E \ll
M_{h^0}$ is necessary, which is in conflict with data. Thus, the
$\mu\to e\gamma$ constraint rules out the Cheng--Sher ansatz with
4th generation lepton in the loop.

Even if we neglect the above shortfall, i.e. if we assume $e$
decouples from $Eh^0$, we still face the constraint on ${\cal
B}(\tau\to\mu\gamma)$, i.e.
 \begin{eqnarray}
 \hspace{-0.5cm}&\,&{\cal B}^{\mathrm{2HDM-III}}(\tau\to \mu\gamma)\nonumber\\
 \hspace{-0.5cm}&=&{\cal B}(\tau\to\mu\bar{\nu}_\mu\nu_\tau)\frac{3\alpha}{2\pi}\frac{m_\mu}{m_\tau}
 |F_{h^0}(x)|^2\nonumber\\
 \hspace{-0.5cm}&=&3.6\times10^{-5} |F_{h^0}(x)|^2.\label{tmg2HDMIII}
 \end{eqnarray}
If we hold that $m_E/M_{h^0} > 0.1$ as reasonable, then $\tau \to
\mu\gamma$ is again ruled out by Eq.~(\ref{tautomugam}) already.

If the 4th generation is found, it seems that the Cheng-Sher
ansatz can not hold for the lepton sector.

\subsection{\boldmath $E E Z'$ Loop Contribution}

For completeness, we discuss $EEZ'$ contribution. $Z'$ is the new
gauge boson associated with an additional Abelian gauge symmetry
$U'(1)$ \cite{PM:2000}. Because the typical constraint on the
$Z$--$Z'$ mixing angel $\theta$ is $\theta <
\mathcal{O}(10^{-3})$, we assume for simplicity that there is no
mixing between $Z'$ and $Z$, i.e. they are also the mass
eigenstates. In terms of physical fields, the Lagrangian
associated with the $U'(1)$ gauge symmetry in the charged lepton
sector is written as
 \begin{eqnarray}
 \mathcal{L}_{Z'}= g_{Z'}\overline{e}_a \gamma^{\mu}
 \left[\epsilon^L_{ab}\,L+\epsilon^R_{ab}\,R
 \right] e_b\,Z'_\mu,
 \end{eqnarray}
where $\epsilon^L_{ab}$ and $\epsilon^R_{ab}$ are the $4\times4$
chiral coupling matrices of $Z'$ with charged leptons
$a,b=1,...,4$ are flavor indices, and ${L(R)}$ is the
left(right)-handed projection operator $(1\mp\gamma^5)/2$. Because
of the reality of the Lagrangian, $\epsilon^{L(R)}$ must be
hermitian.

\begin{figure}[t!]
  \centering
  \includegraphics[height=4.5cm]{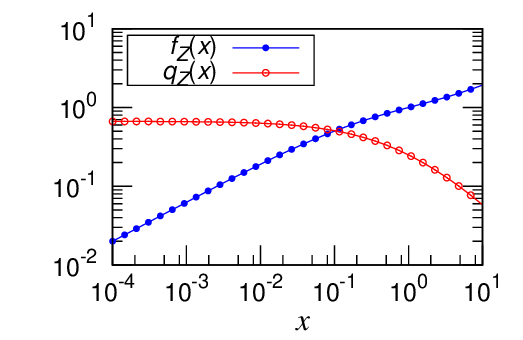}
  \caption{
  Loop functions $f_Z(x)$ and $q_Z(x)$ of Eq.~(\ref{EEZ'}),
  vs $x=m_E^2/M_{Z'}^2$.
  }
  \label{fig:49}
 \end{figure}

After some calculation, we get the dominant contribution to
$g_{\mu} - 2$ from $E E Z'$ loop diagram,
 \begin{eqnarray}
 a_\mu(EEZ')
 &=& \frac{g_{Z'}^2}{8\pi^2}
   \Bigl[x_\mu^{1/2}\,\mathrm{Re}
    \left(\epsilon^L_{\mu E}\epsilon^R_{E\mu}\right) f_Z(x) \nonumber\\
 &\,& \ \ -x_\mu\left(|\epsilon^L_{\mu E}|^2
 +|\epsilon^R_{\mu E}|^2\right) q_Z(x)\Bigr]\label{EEZ'},
  \end{eqnarray}
where $x=m_E^2/M_{Z'}^2$ and $x_\mu=m_\mu^2/M_{Z'}^2$. The loop
functions are
 \begin{eqnarray}
 f_Z(x)&=&\int_{0}^{1}
 du \frac{4u(1-u)x^{1/2}+u^2 x^{3/2}}{u\,x+(1-u)\,}\nonumber\\
 &=&-\frac{3x^{3/2}\log\,x}{(x-1)^3}+\frac{x^{1/2}(x^2+x+4)}{2(x-1)^2},\label{fzx}
 \\
 q_Z(x)&=&\int_0^1du\,\frac{u(1-u)(2-u)}{(1-u)+u\,x}\nonumber\\
 &=&-\frac{x(2x-1)\log x}{(x-1)^4}+\frac{5x^2+5x-4}{6(x-1)^3},\label{qzx}
 \end{eqnarray}
which are plotted in Fig.~\ref{fig:49}.

As a check, we take the case of $\mu\mu Z$. Then
$g_Z=e/\sin\theta_W\cos\theta_W$,
$\epsilon^L_{\mu\mu}=1/2-\sin^2\theta_W$ and
$\epsilon^R_{\mu\mu}$$=-\sin^2\theta_W$. From Eqs.~(\ref{fzx}) and
(\ref{qzx}) we have $f_Z(x_\mu)=2\sqrt{x_\mu}$ and
$q_Z(x_\mu)=2/3$. Hence, we find $a_\mu(\mu\mu Z)
=-(G_F\,m_\mu^2/6\sqrt{2}\pi^2)(1
+2\sin^2\theta_W-4\sin^4\theta_W)$, which is consistent with
Ref.~\cite{Marciano}, as mentioned in Sec.~II-A.

On the other hand, we can also get the contribution to the branching ratio of $\mu\to e\gamma$ from $E E Z'$ loop diagram,
 \begin{eqnarray}
 {\cal B}^{Z'}(\mu\to e\gamma)
 &=& \frac{3 \alpha g_{Z'}^4}{8\pi\,G^2_F\,m^2_\mu\,M_{Z'}^2}|f_Z(x)|^2
     \nonumber\\
 &\,& \ \ \times \Bigl[\,|\epsilon^R_{e E} \epsilon^L_{E \mu}|^2
 +|\epsilon^L_{e E} \epsilon^R_{E \mu}|^2\Bigr]\label{B1}.
  \end{eqnarray}

For $\tau\tau Z'$ in the loop, we find
 \begin{eqnarray}
 a_\mu(\tau\tau Z')
 = \frac{g_{Z'}^2}{4\pi^2} \frac{m_\mu m_\tau}{M_{Z'}^2}
   \mathrm{Re}\left(\epsilon^L_{\mu\tau}\epsilon^R_{\tau\mu}\right),
  \end{eqnarray}
which is given in Ref.~\cite{Chiang:2006we}, but with a sign
error. A similar formula holds for $E$ in the loop. It can be seen
that if $\epsilon^L_{\mu E}$ or $\epsilon^R_{\mu E}$ is zero
(purely right-handed or purely left-handed), $a_{\mu}(EEZ')$ will
become insignificant. This is because a spin flip is required.

For sake of illustration, we take $M_{Z'}=1\,\mathrm{TeV}$,
$m_{E}=250\,\mathrm{GeV}$, $g_{Z'}=0.105$ (predicted from a string
model \cite{Cleaver:1998gc}), and denote $\epsilon^L_{\mu
E}=\epsilon^R_{\mu E} \equiv \epsilon_{\mu E}$ and real, then
  \begin{eqnarray}
 a_\mu(EEZ')
 \sim620\times10^{-11}\,\epsilon_{\mu E}^2\label{EEZ'2}.
  \end{eqnarray}
Furthermore, we assume $\epsilon_{eE}^L=\epsilon_{eE}^R \equiv
\epsilon_{e E}$ and real, then
 \begin{eqnarray}
 {\cal B}^{Z'}(\mu\to e\gamma)
 &\sim&2.47\times10^{-2}\,\,\epsilon_{e E}^2 \, \epsilon_{\mu E}^2.
 \label{BRZp}
 \end{eqnarray}
Comparing Eq. (\ref{difference}) with Eq.~(\ref{EEZ'2}), we need
$\epsilon_{\mu E}=\mathcal{O}(1)$ for $Z^\prime$ to be the main
contributor to muon $g-2$. This is reasonable for a gauge
interaction, as already stated. But one would need a model for why
$\epsilon_{\mu E}$ is of order one. Furthermore, comparing
Eqs.~(\ref{mutoegam}) and (\ref{BRZp}), we find $\epsilon_{e
E}/\epsilon_{\mu E} = \mathcal{O}(10^{-4})$ to satisfy $\mu\to
e\gamma$ constraint. Hence, the model would not only need to
account for $\epsilon_{\mu E} \sim 1$, but $\epsilon_{e E} \ll 1$
as well. If this $Z^\prime$ is a main contributor to muon $g-2$,
it better not couple to electrons.

For completeness, we discuss the contribution to the branching ratio of $\tau\to \mu\gamma$ from $E E Z'$ loop diagram.
Using a similar formula to Eq. (\ref{B1}), we find
 \begin{eqnarray}
 &\,& {\cal B}^{Z'}(\tau\to \mu\gamma) \nonumber\\
 &=& {\cal B}(\tau\to\mu\bar{\nu}_\mu\nu_\tau)\frac{3 \alpha g_{Z'}^4}{8\pi\,G^2_F\,m^2_\tau\,M_{Z'}^2}
     |f_Z(x)|^2 \nonumber\\
 &\,& \ \ \times \Bigl[\,|\epsilon^R_{\mu E} \epsilon^L_{E \tau}|^2
 +|\epsilon^L_{\mu E} \epsilon^R_{E \tau}|^2\Bigr]\label{B2}.
  \end{eqnarray}
If we take the same assumption  and denote $\epsilon^L_{\tau
E}=\epsilon^R_{\tau E} \equiv \epsilon_{\tau E}$ and real, then
 \begin{eqnarray}
 {\cal B}^{Z'}(\tau\to \mu\gamma)
 &\sim&1.52\times10^{-5}\,\,\epsilon_{\mu E}^2 \, \epsilon_{\tau E}^2.
 \label{BRZp2}
 \end{eqnarray}
Comparing Eq. (\ref{tautomugam}) with Eq. (\ref{BRZp2}), taking
$\epsilon_{\mu E} \sim 1$ as before, we find $\epsilon_{\tau E}
\sim 10^{-2}$ is needed to satisfy $\tau\to \mu\gamma$ constraint.
Thus, it does not seem likely that the $Z^\prime$ is the dominant
source for muon $g-2$.

 %

 %
 %

\section{Comparison with MSSM}

From previous discussions, we understand that the loop function
(Inami--Lim functions) does not receive significant enhancement
with heavy particle in the loop, so the coupling strengths become
the crucial factors. For example, the magnitudes of electroweak
contribution $a_\mu(W^+W^-\nu_\mu)$ and $a_\mu(\mu\mu Z)$ both
enter the interesting range $295(88)\times10^{-11}$ because they
just involve gauge couplings, without any off-diagonal
suppression. On the other hand, extreme smallness of ${\cal
B}(\mu\to e\gamma)$ is predicted in the framework of SM, since
there is no tree level FCNC, while loop effects are highly GIM
suppressed by neutrino mass.Are {\it other} sizable electroweak
contribution possible\,? As we mentioned in Sec.~II-B, there is
such a mechanism~\cite{Marciano,Moroi,Everett} in the MSSM.

Simply put, MSSM doubles the number of diagrams of SM. The
corresponding loops to $W^+W^-\nu_\mu$ and $\mu\mu Z$ are
$\text{chargino-chargino}$-$\tilde{\nu}_\mu$ and
$\tilde{\mu}$-$\tilde{\mu}$-$\text{neutralino}$ respectively.
Assuming mass degeneracy of superparticles,
$m_{\text{Higgsino}}=m_{\text{Wino}}=M_{\tilde{\nu}_\mu}=M_{\text{SUSY}}$,
and in the large $\tan\beta$ limit (to compensate the extra
heaviness of $M_{\text{SUSY}}$) \cite{Moroi}, one can get a
sufficient contribution.

However, Ref.~\cite{Marciano} used a different degeneracy
condition, $m_{\text{chargino}} = M_{\tilde{\nu}_\mu} =
M_{\text{SUSY}}$, which would inadvertently send the chargino loop
into a suppression region. Because it is a little subtle, we take
a closer look.

Following the formulas in Ref. \cite{Moroi} and under the
condition
$m_{\text{chargino}}=M_{\tilde{\nu}_\mu}=M_{\text{SUSY}}$, we have
 \begin{eqnarray}
 \sum_i a_\mu(\chi^+_i\chi^-_i\tilde{\nu}_\mu)
 &=& \frac{1}{48\pi^2}\sum_i\left[x_\mu(C^{L*}_iC^L_i+C^{R*}_iC^R_i)
     \right. \nonumber\\
 &&\hspace{1.2cm} \left. - 9x_\mu^{1/2}\text{Re}(C^{L*}_iC^{R}_i)\right],
 \end{eqnarray}
with $x_\mu=m_\mu^2/M_{\text{SUSY}}^2$,
$C^L_i=(\sqrt{2}m_\mu/v\cos\beta)(U_{\chi^-})_{2i}$ and
$C^R_i=-g_2(U_{\chi^+})_{1i}$, while $U_{\chi^+}$ and $U_{\chi^-}$
are unitary matrices and related by
  \begin{eqnarray}
 \left(
   \begin{array}{cc}
     -M_{G_2} & \frac{g_2v\cos\beta}{\sqrt{2}} \\
     -\frac{g_2v\sin\beta}{\sqrt{2}} & \mu_H \\
   \end{array}
 \right)=M_{\text{SUSY}}\,U_{\chi^+}U_{\chi^-}^\dagger.
 \end{eqnarray}
After some calculation, we get
 \begin{eqnarray}
 \sum_i a_\mu(\chi^+_i\chi^-_i\tilde{\nu}_\mu)
 \sim837\times10^{-5}x_\mu.
 \end{eqnarray}
With $x_\mu=m_\mu^2/M_{\text{SUSY}}^2 \lesssim (10^{-7})$
typically, the chargino-chargino-$\tilde{\nu}_\mu$ loop
contribution is subdominant.

\section{Summary}

In this paper, we consider the existence of 4th generation leptons
and discussed their impact on $a_\mu$.
In the SM, 2HDM-I and II, the 4th generation seems irrelevant to
the $\Delta a_\mu$ puzzle because of the smallness of
$|V_{N\mu}|$. However, this off-diagonal factor also protects
these models from the stringent $\mu\to e \gamma$ and
$\tau\to\mu\gamma$ constraints.
In the 2HDM-III, applying the Cheng--Sher ansatz with 4th
generation to charged leptons, one has a strong conflict with
${\cal B}(\mu\to e\gamma)$ and even ${\cal B}(\tau\to \mu\gamma)$.
Hence, if 4th generation is found, the Cheng--Sher ansatz cannot
hold for lepton sector. This may be reasonable since the lepton
mixing pattern seems different from quarks.

Our analysis illustrates why the well known SUSY mechanism is
favored. Enhancement to $a_\mu$ and suppression to ${\cal
B}(\mu\to e\gamma)$ in the MSSM both bear similarities to the SM.
It is interesting that the required large $\tan\beta$ enhancement
for the SUSY effect renders the {\it negative} contribution from
$H^+H^-N$ negligible, hence MSSM and 4th generation can coexist.

\vskip 0.2cm \noindent{\bf Acknowledgement}.\ This work is
supported in part by the National Science Council of R.O.C. under
grant numbers NSC96-2739-M-002-005, NSC96-2811-M-002-042, and
NSC96-2811-M-002-097. We thank C.~Biggio, G.~Burdman, C.W.~Chiang,
and A.~Czarnecki for communications.

\end{document}